# The Persistent Qubit

Isaac L. Chuang *and* Y. Yamamoto

[1]*ERATO Quantum Fluctuation Project*
*Edward L. Ginzton Laboratory, Stanford University, Stanford, CA 94305*
(April 25, 1996)

The construction of large, coherent quantum systems necessary for quantum computation remains an entreating but elusive goal, due to the ubiquitous nature of decoherence. Recent progress in quantum error correction schemes have given new hope to this field, but thus far, the codes presented in the literature assume a restricted number of errors and error free encoding, decoding, and measurement. We investigate a specific scenario without these assumptions; in particular, we evaluate a scheme to preserve a single quantum bit against phase damping using a three-qubit encoding based on Shor. By applying a new formalism which gives simple operators for decoherence and noisy logic gates, we find the fidelity of the stored qubit as a function of time, including decoherence which occurs not only during storage but also during processing. We generalize our results to include any source of error, and derive an upper limit on the allowable decoherence per timestep. Physically, our results suggest the feasibility of engineering artificial metastable states through repeated error correction.

42.50.Ar,89.80.th,42.79.Ta,03.65.Bz

## I. INTRODUCTION

Technological progress has increasingly enabled us to fabricate and manipulate small quantum mechanical systems in which coherence is preserved. For example, the development of cold atom and ion traps, and coherent quantum dot devices has inspired the possibility that soon, implementation of "designer" quantum systems may be feasible. These will be useful for fundamental investigations of physics, in particular, by addressing open questions such as the transition from quantum to classical behavior. They may also allow utilization of the superposition and non-locality properties of quantum mechanics for information processing, as in quantum cryptography [1] and computing [2].

Quantum computation has recently attracted a great deal of attention as a result of new efficient algorithms [3]. This has been accompanied experimental results [4,5] which give hopeful signs that it might indeed be possible in the future to rapidly factor large numbers using such machines. However the present technology is still rather crude, and even factoring a number like 15 will be a tour de force. In an ion trap model it would take of the order of 25,000 laser pulses to achieve this task, and this assumes perfect operation of the computer. Although the best way not to have to recover from errors is to avoid them, it seems unlikely that computation with so many laser pulses can be error-free.

Classically, this problem is rectified by error correction schemes, but these techniques do not apply in general to quantum bit (qubit) errors, because they do not preserve quantum coherence; direct measurement of a qubit will destroy its coherence. Furthermore, a quantum superposition state is fundamentally difficult to maintain, due to unwanted environmental interactions which lead to *decoherence* [6]. What is needed is some way to correct a qubit state without ever completely measuring it. Fortunately, such schemes for quantum error correction have recently been developed, and they allow certain independent errors may be corrected. Nine [7], seven [8], and five [9,10] qubit codes have been discovered which perfectly correct single qubit errors of any kind. With these successes comes the hope that construction of small, coherent, quantum systems may be possible despite decoherence.

However, an important issue that remains to be addressed is the effect of errors which occur *during processing* required by the error correction scheme. All of the theoretical quantum error correction results presented in the literature so far assume perfect operation of the coding, measurement, and decoding circuits. As various studies of the effects of decoherence on quantum computers have shown [11–15], this is not a realistic assumption when the time scale for decoherence is comparable to the coding and decoding time of the circuit, as is the case in current experimental systems. To address this issue, we present here a systematic analysis of the impact of imperfect processing on a model quantum error correction system.

Specifically, we analyze a three-bit code which perfectly corrects for any single qubit error due to a specific kind of decoherence known as phase damping. We apply this code to a system in which periodic correction is applied to artificially lengthen the lifetime of an encoded qubit state, which we refer to as a "persistent qubit." We perform numerical simulations which include the effects of decoherence during the logic operations, and calculate the fidelity of the persistent qubit a function of the decoherence per timestep figure of merit $\lambda$. Our results indicate the existence of an upper limit $\lambda_{crit}$ which must be achieved in order to gain any advantage from error correction.

These results are made possible in part by the development of a mathematical model based on linear operators which gives concise descriptions of the effects of decoherence. From this theory, we construct operators for phase and amplitude damping, and for noisy logic gates. This paper is thus organized as follows: in the first part, we present our theoretical model of decoherence. These results are applied to an analysis of the persistent qubit model in the second half of the paper. We conclude with a discussion of experimental possibilities.

## II. THEORY: $A_K$ MODEL OF DECOHERENCE

We begin by presenting an unusual mathematical theory for describing decoherence, based on linear operators. Historically, the linear operator formalism has received little attention in the community [16,17], but it is particularly well suited to for manipulating the finite Hilbert spaces of quantum bits. Furthermore, it is fundamentally equivalent to the usual density matrix approach, but motivates an alternative interpretation based on the evolution of a *single wavefunction*. We find concise descriptions for decoherence operators of two kinds (phase and amplitude damping), and also introduce the notion of a *noisy logic gate operator*. The results are used to calculate the fidelity of noisy rotation and controlled-not operators.

### A. Decoherence – Density Matrices

Decoherence occurs due to unwanted interactions between our quantum system and its environment. These interactions cause information to leak out of the system, and fluctuations to enter it. Typically, this process is described by density matrices; for example, the pure state

$$|\psi\rangle = a\,|0\rangle + b\,|1\rangle\,, \tag{1}$$

written in the basis of energy eigenstates (which we shall use as the "computational basis" later) has the density matrix

$$\rho_{in} = \begin{bmatrix} |a|^2 & ab^* \\ a^*b & |b|^2 \end{bmatrix}, \tag{2}$$

where the diagonal elements give the probabilities of finding the system in the zero and one states, and the nonzero off-diagonals connote the existence of some coherence. The signature of loss of quantum coherence is decay of off-diagonal elements,

$$\rho_{out} = \begin{bmatrix} |a|^2 & ab^*e^{-\lambda} \\ a^*be^{-\lambda} & |b|^2 \end{bmatrix}. \tag{3}$$

For example, this process may occur when a single photon qubit is transmitted through a fiber whose length is randomly modulated by acoustic waves, introducing *phase damping* – the fluctuations cause uncertainty in the arrival time and thus destroy information in the variable conjugate to the amplitude. The average effect after many phase kicks is a damping process, whose net effect is the reduction in the *fidelity* $\mathcal{F}$ of the received qubit,

$$\mathcal{F}(\psi) = \min_\psi \left[ \langle \psi | \rho_{out} | \psi \rangle \right] \tag{4}$$

$$= \min_{a,b} \left[ 1 + 2|a|^2|b|^2(e^{-\lambda} - 1) \right] \tag{5}$$

$$\geq \frac{1 + e^{-\lambda}}{2} \approx 1 - \frac{\lambda}{2}\,. \tag{6}$$

Note that (following Schumacher [18]) we define $\mathcal{F}$ as the *minimum* value of the overlap between the initial and final wavefunctions, because in general, a quantum computer may access all states in the Hilbert space of $|\psi\rangle$.

In general, decoherence may introduce effects other than just decay of the off diagonal terms. Relaxation processes (otherwise known as *amplitude damping*) cause energy to be lost from the system as well as phase information. The most general description is given by interacting the system unitarily with some initial state $|e\rangle$ of the environment then tracing over the environment to get the final state, i.e.

$$\rho_{out} = \text{Tr}_{env} \left[ U(\rho_{in} \otimes |e\rangle\langle e|)U^\dagger \right]. \tag{7}$$

Mathematically, we may introduce a complete set of states $|\mu_k\rangle$ for the environment,

$$\sum_k |\mu_k\rangle\langle\mu_k| = I \tag{8}$$

such that we may express the final state of the system as

$$\rho_{out} = \sum_k A_k \rho_{in} A_k^\dagger\,, \tag{9}$$

where $A_k$ are linear operators (not necessarily Hermitian) in the Hilbert space of the system, given by

$$A_k = \langle \mu_k | U | e \rangle\,. \tag{10}$$

Note that by unitarity of $U$, we must have that $\sum_k A_k^\dagger A_k = \mathcal{I}$, and in this sense, the $A_k$ describe all the possible processes which may happen to the system due to decoherence.

For example, the two operators

$$A_0 = \begin{bmatrix} 1 & 0 \\ 0 & e^{-\lambda} \end{bmatrix} \tag{11}$$

$$A_1 = \begin{bmatrix} 0 & 0 \\ 0 & \sqrt{1 - e^{-2\lambda}} \end{bmatrix} \tag{12}$$

give us

$$\rho_{out} = \sum_k A_k \begin{bmatrix} |a|^2 & ab^* \\ a^*b & |b|^2 \end{bmatrix} A_k^\dagger \qquad (13)$$

$$= \begin{bmatrix} |a|^2 & ab^*e^{-\lambda} \\ a^*be^{-\lambda} & |b|^2 e^{-2\lambda} \end{bmatrix} + \begin{bmatrix} 0 & 0 \\ 0 & |b|^2(1-e^{-2\lambda}) \end{bmatrix} \qquad (14)$$

$$= \begin{bmatrix} |a|^2 & ab^*e^{-\lambda} \\ a^*be^{-\lambda} & |b|^2 \end{bmatrix}. \qquad (15)$$

Thus, $A_0$ and $A_1$ describe phase damping! At this point we have simply pulled these expressions out of thin air; however, there is actually good physical motivation for these choices – this is the subject of the next section. Furthermore, it is interesting to note that this choice of operators is not unique; there exist different $A_k$ which also describe phase damping (mathematically, they correspond to a change of basis for the environment states $|\mu_k\rangle$ in Eq.(10)). In addition, other $A_k$ may describe different forms of decoherence such as amplitude damping.

### B. The Single Wavefunction Model

Although the description of decoherence using density matrices and the $A_k$ operators of Eq.(9) is completely general, unfortunately, some important information about the structure of the decoherence is hidden by the density matrix formalism (for example, what is the minimum number of pure states which a mixture $\rho$ can be decomposed into?). In particular, we may view the $A_k$ as operators which divide the Hilbert space of the final state into different partitions indexed by $k$. By summing over $k$ to get $\rho_{out}$, we loose this information about the partitioning.

This useful information can be preserved by withholding the sum over $k$ performed in Eq.(9), and keeping track of the evolution of each $A_k \rho A_k^\dagger$ separately. An equivalent technique which simplifies the bookkeeping is the *single wavefunction* model, in which mixed states are written as direct sums of pure states. For example, the mixed state

$$|\psi_{out}\rangle = \bigoplus_k |\phi_k\rangle \qquad (16)$$

could result from a measurement of $|\psi_{in}\rangle$, where $p_k = \langle \phi_k|\phi_k\rangle$ is the probability of obtaining the observable eigenstate $|\phi_k\rangle/\sqrt{p_k}$. Thus, expectation values are defined as

$$\langle \psi_{out}|O|\psi_{out}\rangle = \sum_k \langle \phi_k|O|\phi_k\rangle, \qquad (17)$$

since the different $|\phi_k\rangle$ live in completely separate spaces. In this language, Eq.(9) may be re-expressed as

$$|\psi_{out}\rangle = \bigoplus_k A_k|\psi_{in}\rangle, \qquad (18)$$

because

$$\rho_{out} = |\psi_{out}\rangle\langle\psi_{out}| = \sum_k A_k|\psi_{in}\rangle\langle\psi_{in}|A_k^\dagger. \qquad (19)$$

For example, if we use the $A_k$ defined in Eqs(11-12), we obtain for $|\psi_{in}\rangle = a|0\rangle + b|1\rangle$ the output mixed state

$$|\psi_{out}\rangle = \left[a|0\rangle + be^{-\lambda}|1\rangle\right] \oplus \left[b\sqrt{1-e^{-2\lambda}}|1\rangle\right] \qquad (20)$$

Physically we may understand these two states as resulting from an implicit ("non-referred") indirect ("POVM") measurement [19] of the system in which the environment acts as a probe. It is implicit because in reality, no observer ever refers to the measurement result – that is left unknown. In particular, the interaction may be modeled as shown in Fig. 1, where the environment is an interferometer containing one excitation, and cross-phase modulation (XPM) via the Hamiltonian

$$H_I = \chi a^\dagger a e_a^\dagger e_a \qquad (21)$$

with our qubit (the system) for time $\tau = (\cos^{-1} e^{-\lambda})/\chi$ causes the interferometer to become partially unbalanced. The unitary operator may be understood as transforming

$$|0\rangle|01\rangle \rightarrow |0\rangle|01\rangle \qquad (22)$$

$$|1\rangle|01\rangle \rightarrow e^{-\lambda}|1\rangle|01\rangle + \sqrt{1-e^{-2\lambda}}|1\rangle|10\rangle, \qquad (23)$$

where the first label denotes the system and the second, the environment. An implicit measurement of the interferometer's output occurs because the system leaves the environment behind – this measurement is equivalent to tracing over the environment degrees of freedom. A measurement result of $|01\rangle$ (no unbalancing) corresponds to the first bracketed term in Eq.(20), and a result of $|10\rangle$ (unbalanced) to the second term. Note that the density matrix for Eq.(20) is exactly the same as Eq.(15), so that for an ensemble of states, the net effect is phase damping.

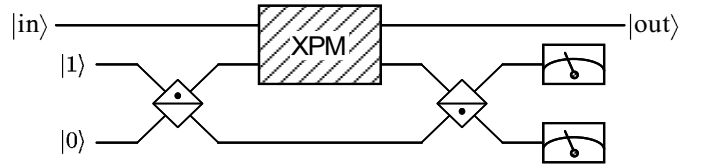

FIG. 1. Physical model of interaction with environment leading to phase damping. The two 50/50 beamsplitters are inverses of each other. The meters indicate an implicit measurement of the state.

### C. The Phase Damping Operator

We shall now see how the single wavefunction model of decoherence can be useful. Suppose that before the environment is measured, a unitary transform $U$ is performed as shown in Fig. 2. In particular, let us choose the orthonormal basis

$$|\mu_0\rangle = \frac{(1+e^{-\lambda})|01\rangle + \sqrt{1-e^{-2\lambda}}|10\rangle}{\sqrt{2(1+e^{-\lambda})}} \quad (24)$$

$$|\mu_1\rangle = \frac{(1-e^{-\lambda})|01\rangle - \sqrt{1-e^{-2\lambda}}|10\rangle}{\sqrt{2(1+e^{-\lambda})}} \quad (25)$$

such that the unitary transform of Eqs.(22-23) becomes

$$|0\rangle|e\rangle \to \sqrt{\alpha}|0\rangle|\mu_0\rangle + \sqrt{1-\alpha}|0\rangle|\mu_1\rangle \quad (26)$$
$$|1\rangle|e\rangle \to \sqrt{\alpha}|1\rangle|\mu_0\rangle - \sqrt{1-\alpha}|1\rangle|\mu_1\rangle, \quad (27)$$

where $\alpha = (1+e^{-\lambda})/2$, and

$$|e\rangle = \sqrt{\alpha}|\mu_0\rangle + \sqrt{1-\alpha}|\mu_1\rangle \quad (28)$$

is the initial state of the environment. The two possible measurement results now become $|\mu_0\rangle$ and $|\mu_1\rangle$, so that the final state of the system may be written as

$$|\psi_{out}\rangle = \sqrt{\alpha}\left[a|0\rangle + b|1\rangle\right] \oplus \sqrt{1-\alpha}\left[a|0\rangle - b|1\rangle\right] \quad (29)$$

where the two terms result from obtaining $|\mu_0\rangle$ and $|\mu_1\rangle$ respectively. Since $|\psi_{out}\rangle\langle\psi_{out}|$ gives the same density matrix as in Eq.(15), this process is also a statistically valid single wavefunction description of phase damping.

Of course, in real distributed decoherence processes involving many modes of the environment, there will be no basis rotation $U$ or implicit measurement of $|\mu_0\rangle$ or $|\mu_1\rangle$. Rather, the point is that Eqs.(26-27) give an *equivalent model* which can be used to describe all phase damping processes occurring to a single qubit. Because experimental observations of quantum systems are always stochastic, there is no observational difference between this model and any other model of phase damping. However, a simple model which distills the essence of the process can be a powerful tool for understanding the physics. In particular, the value of this model is the elegance of the following mathematical conclusion: *in the single wavefunction picture, we may say that phase damping either leaves the bit alone, or causes the phase of the bit to be flipped.* We shall see in Sec. III A how this helps in devising a scheme for correcting errors due to decoherence.

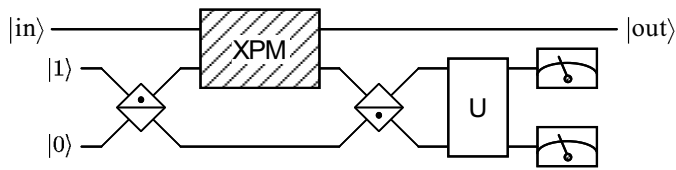

FIG. 2. Implicit measurement of the environment performed in a different basis.

### D. Decoherence of $N$ qubits

We now show how simple operators can be constructed which describe the effect of phase damping on a set of $N$ qubits. The essential idea is that by assuming that the environment acts independently on each qubit, we may find a product form for the phase damping operator $A_k$. First, note that the unitary transform which entangles a single qubit and the environment given in Eqs.(26-27) can be rewritten using a different basis for the qubit. Suppose we choose the "Bell basis" states

$$|+\rangle = \frac{|0\rangle + |1\rangle}{\sqrt{2}} \quad (30)$$

$$|-\rangle = \frac{|0\rangle - |1\rangle}{\sqrt{2}}, \quad (31)$$

then we have that

$$|+\rangle|e\rangle \to \sqrt{\alpha}|+\rangle|\mu_0\rangle + \sqrt{1-\alpha}|-\rangle|\mu_1\rangle \quad (32)$$
$$|-\rangle|e\rangle \to \sqrt{\alpha}|-\rangle|\mu_0\rangle + \sqrt{1-\alpha}|+\rangle|\mu_1\rangle, \quad (33)$$

so that an initial state

$$|\psi'_{in}\rangle = a|+\rangle + b|-\rangle \quad (34)$$

decoheres into the final mixed state

$$|\psi'_{out}\rangle = \sqrt{\alpha}\left[a|+\rangle + b|-\rangle\right] \oplus \sqrt{1-\alpha}\left[a|-\rangle + b|+\rangle\right] \quad (35)$$

such that we may say phase damping causes bit flip errors when the Bell basis is used as the representation (or "computational") basis for a qubit. Note that the probability of that the single bit is flipped is $1-\alpha$. In terms of Eq.(18), the projection operators are

$$A_0^{pd} = \sqrt{\alpha}\begin{bmatrix} 1 & 0 \\ 0 & 1 \end{bmatrix} \quad (36)$$

$$A_1^{pd} = \sqrt{1-\alpha}\begin{bmatrix} 0 & 1 \\ 1 & 0 \end{bmatrix}. \quad (37)$$

$A_1^{pd}$ is proportional to the Pauli matrix $\sigma_x$, which performs a spin flip (in the $\{|0\rangle, |1\rangle\}$ basis we would have $\sigma_z$ instead). From this viewpoint, the bit-flip interpretation of phase damping is manifestly clear.

Let us extend this analysis to a system of two qubits ($a$ and $b$) interacting with independent reservoirs ($e_a$ and $e_b$). For example, the interaction Hamiltonian could be written in spin notation as

$$H_{I_2} = \frac{\chi}{2}(1-\sigma_z^a)\sigma_y^{e_a} + \frac{\chi}{2}(1-\sigma_z^b)\sigma_y^{e_b}, \quad (38)$$

using Pauli matrices (when $a$ has spin down (up) then $(1-\sigma_z^a)/2$ evaluates to zero (one)). Because only a single excitation is contained in the interferometer of Fig. 1, we may model phase damping as a *controlled rotation* of one environmental mode by each qubit.

Unitary evolution via

$$U = \exp\left[\frac{i\cos^{-1}(2\alpha-1)}{\chi}H_{I_2}\right] \quad (39)$$

entangles the qubits with the environment, and tracing out the environment in the $|\mu_i\rangle$ basis gives the projection operators, calculated from Eq.(10):

$$\tilde{A}_0 = \alpha\,(I\otimes I) \quad (40)$$
$$\tilde{A}_1 = \sqrt{\alpha(1-\alpha)}\,(I\otimes\sigma_x) \quad (41)$$
$$\tilde{A}_2 = \sqrt{\alpha(1-\alpha)}\,(\sigma_x\otimes I) \quad (42)$$
$$\tilde{A}_3 = (1-\alpha)\,(\sigma_x\otimes\sigma_x), \quad (43)$$

where the two operators on either side of $\otimes$ act on $a$ and $b$ respectively. Generalizing from this, it may be shown that the projection operators for phase damping of $N$ qubits is

$$\tilde{A}_k = \sqrt{\alpha^{N-h(k)}(1-\alpha)^{h(k)}}\bigotimes_{n=0}^{N-1}(\sigma_x)^{h(2^n\wedge k)}, \quad (44)$$

where $h(k)$ is the number of one's in the binary bit-string form of $k$ (i.e., the Hamming weight), and $\wedge$ is the boolean AND operator. $k$ ranges from 0 to $2^N - 1$. The effect of the exponent $h(2^n\wedge k)$ is to select either $I$ or $\sigma_x$ for the $n^{th}$ qubit based on whether the $n^{th}$ bit in $k$ is zero or one. The $\alpha^{N-h(k)}(1-\alpha)^{h(k)}$ prefactor gives the probability of each projection, from which it is evident that multiple bitflips are less likely than few flips. From this calculation, we conclude that *with no approximation*, the effect of phase damping on a set of $N$ qubits can be described by projections into $2^N$ possible states in which different bits are flipped, and the probability of having $m$ bits flipped is

$$\text{Prob}(m) = \binom{N}{m}\alpha^{N-m}(1-\alpha)^m. \quad (45)$$

Note that $\sum_{m=0}^{N}\text{Prob}(m) = 1$ as expected. Furthermore, the mixed state resulting from phase damping of any state $|\psi\rangle$ can be immediately calculated using the above result; let $b$ denote the bit string formed by $N$ qubits, i.e., $|b\rangle = |b_{N-1}\cdots b_1 b_0\rangle$. If the input state is

$$|\psi_{in}\rangle = \sum_b c_b |b\rangle, \quad (46)$$

then the mixture resulting from phase damping is

$$|\psi_{out}\rangle = \bigoplus_k \tilde{A}_k |\psi_{in}\rangle \quad (47)$$
$$= \bigoplus_k \sqrt{\alpha^{N-h(k)}(1-\alpha)^{h(k)}}\sum_b c_b |b\text{ XOR }k\rangle, \quad (48)$$

where $b$ XOR $k$ denotes the binary exclusive-or of the two bit strings. This demonstrates explicitly that the effect of phase damping on $N$ qubits is the creation of a mixed state which may be described as a direct sum of states in which bits are flipped according to a Bernoulli process with probability $1 - \alpha$. This result provides us with an efficient computational tool for calculating the effects of decoherence on a register of $N$ qubits, and will be useful in analyzing an imperfect quantum memory in Sec. III C.

### E. Amplitude Damping

Physically, the effect of phase damping may be understood to be an analog of the "$T_2$" spin depolarization effects observed in nuclear magnetic resonance spectroscopy. Of course, one must be careful to distinguish ensemble time-scales from what we are interested in here, the dephasing of a single spin or two spins relative to each other. The analog of "$T_1$" spin-lattice effects, in which energy is lost from a single spin to the environment is *amplitude damping*. This also describes relaxation processes such as spontaneous emission.

A simple operator description of the amplitude damping of a single qubit may also be derived, just as was done above for phase damping. The effect of energy loss to the environment (relaxation) is usually described by a master equation [20] which, in the Born-Markov approximation, results in the density matrix evolution

$$\begin{bmatrix} a & b \\ b^* & c \end{bmatrix} \to \begin{bmatrix} e^{-\gamma_0}a + (1-e^{-\gamma_1})c & be^{-(\gamma_0+\gamma_1)/2} \\ b^*e^{-(\gamma_0+\gamma_1)/2} & (1-e^{-\gamma_0})a + e^{-\gamma_1}c \end{bmatrix} \quad (49)$$

for a single qubit. Equivalently, we may write that

$$|0\rangle|01\rangle \to \sqrt{1-p}\,|0\rangle|01\rangle + \sqrt{p}\,|1\rangle|00\rangle \quad (50)$$
$$|1\rangle|01\rangle \to \sqrt{1-q}\,|1\rangle|01\rangle + \sqrt{q}\,|0\rangle|11\rangle, \quad (51)$$

where $p = 1 - e^{-\gamma_0}$ and $q = 1 - e^{-\gamma_1}$ are the probabilities of upward and downward transitions, respectively. Here, $|01\rangle$ is a convenient choice for the initial state of the environment. When $\gamma_0 = 0$ we have the usual case of damping to a reservoir at $T = 0$, which describes, for example, the scattering of photons out of a single mode fiber. For nonzero $\gamma_0$ and $\gamma_1$, we have the stationary state

$$\frac{1}{p+q}\begin{bmatrix} q & 0 \\ 0 & p \end{bmatrix}, \quad (52)$$

which describes the system after it has come into equilibrium with a reservoir at temperature

$$k_B T = \frac{\Delta E}{\ln\frac{q}{p}}, \quad (53)$$

assuming a Boltzmann distribution of energies, and an energy difference between the $|1\rangle$ and $|0\rangle$ states of $\Delta E$.

From Eqs.(50-51), we may immediately read off the amplitude damping operators

$$A_{00}^{ad} = \sqrt{p}\,|1\rangle\langle 0| \qquad (54)$$
$$A_{01}^{ad} = \sqrt{1-p}\,|0\rangle\langle 0| + \sqrt{1-q}\,|1\rangle\langle 1| \qquad (55)$$
$$A_{11}^{ad} = \sqrt{q}\,|0\rangle\langle 1|\,. \qquad (56)$$

These may be verified immediately by noting that $\sum_k A_k^{ad} \rho A_k^{ad\dagger}$ gives the same result as Eq.(49). Thus, in general, amplitude damping of an initially pure state qubit will result in a mixed state composed of three pure states, described by the three $A_k^{ad}$ operators above.

An interesting analogy can be made between the phase damping and amplitude damping cases in the following case. Consider the amplitude damping to a reservoir at $T=0$ of the state

$$|\psi\rangle = a\,|01\rangle + b\,|10\rangle\,. \qquad (57)$$

This describes the two-mode output of an optical beamsplitter of angle $\tan^{-1}(a/b)$ with a single photon incident into one input port and vacuum into the other. Equal amplitude damping of *both* the output modes by $\gamma$ results in the mixed state output

$$|\psi_{out}\rangle = e^{-\gamma}\left[a\,|01\rangle + b\,|10\rangle\right] \oplus (1-e^{-\gamma})|00\rangle\,. \qquad (58)$$

Physically, this occurs because only one photon ever exists; if it is lost, the final state must be the vacuum. Otherwise, because of the balanced arrangement of the loss, the initial state is left unchanged. We may also describe this by the transformations

$$|01\rangle|00\rangle \rightarrow \sqrt{e^{-\gamma}}\,|01\rangle|00\rangle + \sqrt{1-e^{-\gamma}}\,|00\rangle|01\rangle \qquad (59)$$
$$|10\rangle|00\rangle \rightarrow \sqrt{e^{-\gamma}}\,|10\rangle|00\rangle + \sqrt{1-e^{-\gamma}}\,|00\rangle|10\rangle\,. \qquad (60)$$

Notice that the $|01\rangle$ and $|10\rangle$ final states of the environment leave the system in the same state, $|00\rangle$, with the same probability amplitude. This equivalence means that $A_{01} = A_{10}$ so they may be combined. Letting $A_0 = A_{00}$ and $A_1 = A_{10}$, we find that the damping of the *dual-rail qubit* state [21] of Eq.(57) may be described by the two $A_k$ operators

$$A_0^{dr} = \sqrt{e^{-\gamma}}\,I \qquad (61)$$
$$A_1^{dr} = \sqrt{1-e^{-\gamma}}\left[|00\rangle\langle 01| + |00\rangle\langle 10|\right]\,. \qquad (62)$$

This result is the basis for an optical quantum bit regeneration scheme [22], which uses a kind of quantum nondemolition measurement to detect the jump into the vacuum state described by $A_1$.

Comparing Eqs.(61-62) with the operator description of the phase damping of a single qubit, Eqs.(36-37), we find an interesting similarity: in both cases, one of the operators is proportional to the identity. Thus, we may interpret the decoherence process for these cases as sometimes *completely leaving the wavefunction unchanged*. However, such a conclusion does not seem possible for the amplitude damping of a *single* qubit – no linear combination of Eqs.(54-56) give the identity operator. This suggests that a more complex circuit may be required for correction of errors due to amplitude damping, as compared to those due to phase damping. This is physically reasonable, because the amplitude damping process affects the *diagonal* terms of the density matrix as well as the off-diagonals; it is a combination of the effects of "pure" dephasing and relaxation. In this sense, it is harder to retrieve from the environment information required to reconstruct a state damaged by amplitude damping than by phase damping.

### F. Noisy Logic Gate Operators

So far, we have constructed a mathematical description of the decoherence of "idle" qubits – other than the coupling to the environment, each qubit is assumed to be interacting with nothing. Of great concern is what happens when decoherence occurs during *conditional dynamics* [23], that is, while two or more qubits are interacting. With what probability will the correct result be obtained? Will the impact of these errors be more significant than "memory" errors?

Using a theory of continuous measurement, Pellizzari et. al. [15] have performed simulations to address this issue for a ion-trap quantum computer. They calculated the fidelity of a controlled-not (CN) gate with simultaneous spontaneous emission (amplitude damping) and cavity decay. Here, we present an alternative approach which uses the linear operator and single wavefunction theory described above. We construct *noisy logic operators* which describe the operation of imperfect quantum logic gates, and use these to evaluate the fidelity of the rotation and CN gates.

The effect of simultaneous decoherence and logic may be modeled by introducing additional interactions with extra qubits which model the environment. In general the true Hamiltonian is

$$H_{noisy} = g H_{logic} + \lambda_0 H_{env}\,, \qquad (63)$$

where $g$ and $\lambda_0$ are coupling constants for the logic interaction and decoherence, respectively. Conditional dynamics occurs due to unitary evolution according to

$$U = e^{iH_{noisy}\tau}\,, \qquad (64)$$

where $\tau$ is selected such that $\exp[ig\tau H_{logic}]$ gives the desired interaction. During this same time period, unwanted environmental interactions due to $H_{env}$ cause the system to become entangled with the environment; this leads to *decoherence*, because information from the system is left behind in the environment. The decoherence experienced during a single timestep $\tau$ is $\lambda_0/g$, which we define to be the parameter $\lambda$, known as the *decoherence per timestep* figure of merit [24].

Now, the coupling of the system to the environment is typically through many modes; in other words, the Hamiltonian takes the form $H_{env} = \sum_n q_{sys} e_n$. Because the environment contains a large number of modes, no Jaynes-Cummings type revival ever occurs (and information once lost to the environment never spontaneously comes back to the system). However, we have shown in the previous sections how a single-mode environment can be used to model the effects of decoherence on a single qubit. Thus, we shall use an $H_{env}$ similar to that employed in Eq.(38), namely,

$$H_{env} = \frac{\chi}{2}(1 - \sigma_z^a)\sigma_y^{e_a} \qquad (65)$$

for a single qubit $a$. Note that because our model of the environment is a single mode, Rabi oscillations will occur on a time-scale determined by the coupling constant $\chi$, so in order to properly model decoherence, we must take care to limit ourselves to the first cycle. Thus, the decoherence per timestep parameter is given by

$$\lambda = -\ln|\cos\chi|. \qquad (66)$$

This single-mode environment model is an approximation of reality which shall be utilized in the remainder of the paper. Intuitively, it is acceptable because since after all, the system contains only a single qubit worth of information which can be lost. The detailed relationship between our model and other specific microscopic environmental interactions will be described elsewhere.

The effect of phase damping on a logic gate can be modeled as shown by the following example. The Hamiltonian for the $\pi/4$ rotation gate (useful for transforming between the computational and Bell states) acting on mode $a$ may be written using spin notation as

$$H_R = \frac{\pi}{4}\sigma_y^a + \frac{\chi}{2}(1 - \sigma_z^a)\sigma_y^{e_a}, \qquad (67)$$

where the first term gives the desired rotation

$$R = \exp\left[\frac{i\pi}{4}\sigma_y^a\right] = \frac{1}{\sqrt{2}}\begin{bmatrix} 1 & 1 \\ -1 & 1 \end{bmatrix}, \qquad (68)$$

and the second term describes phase damping. The unitary evolution $U_R = \exp(iH_R)$ simultaneously performs the desired rotation while entangling the qubit with the environment. To obtain the final state of the qubit $a$, the environment is traced out, leaving $a$ in a mixed state. This is exactly the same as in Eq.(7), as described in Sec. II A.

The operators $A_k$ in Eq.(10) model the effect of decoherence on an idle qubit. In a similar manner, we may write operators $R_k$ which describe the simultaneous effect of rotation and phase damping. It is convenient to take the initial state of the environment to be $|0\rangle$ (spin down), such that

$$R_0 = \langle 0|U_R|0\rangle = \frac{1}{2}\left(\exp\begin{bmatrix} 0 & \frac{\pi}{4} \\ -\frac{\pi}{4} & i\chi \end{bmatrix} + h.c.\right) \qquad (69)$$

and

$$R_1 = \langle 1|U_R|0\rangle = \frac{1}{2i}\left(\exp\begin{bmatrix} 0 & \frac{\pi}{4} \\ -\frac{\pi}{4} & i\chi \end{bmatrix} - h.c.\right). \qquad (70)$$

Thus, $|\psi_{out}\rangle = \bigoplus_k R_k|\psi\rangle$ is the output of a rotation gate with phase damping parameterized by $\chi$. Note that when $\chi = 0$, $R_0 = R$ and $R_1 = 0$, which gives the expected result for the ideal case, with no decoherence. The fidelity of the rotation operator is found to be

$$\mathcal{F}_R = \min_\psi \sum_k |\langle\psi|R^\dagger R_k|\psi\rangle|^2 \approx 1 - 0.40\lambda, \qquad (71)$$

to lowest order in $\lambda$.

A similar result can be derived for the CN gate with simultaneous phase damping, using the Hamiltonian

$$H_{CN} = H_{CN}^0 + \frac{\chi}{2}(1-\sigma_z^a)\sigma_y^{e_a} + \frac{\chi}{2}(1-\sigma_z^b)\sigma_y^{e_b}, \qquad (72)$$

where the transform is $U_{CN} = \exp[iH_{CN}]$, the ideal gate has the Hamiltonian

$$H_{CN}^0 = \frac{\pi}{2}(1-\sigma_z^a)(\sigma_x^b - 1) \qquad (73)$$

and the ideal operation is

$$C = \exp\left[iH_{CN}^0\right] = \begin{bmatrix} 1 & 0 & 0 & 0 \\ 0 & 1 & 0 & 0 \\ 0 & 0 & 0 & 1 \\ 0 & 0 & 1 & 0 \end{bmatrix}, \qquad (74)$$

giving the "noisy" CN gate operators is described by the interaction operators

$$C_k = \langle k|U_{CN}|00\rangle \qquad (75)$$

for $k = \{00, 01, 10, 11\}$. Using this formalism, we calculate the fidelity of a CN gate with simultaneous phase damping to be

$$\mathcal{F}_{CN} \approx 1 - 0.86\lambda, \qquad (76)$$

to lowest order in $\lambda$.

Similarly, the effect of amplitude damping of a single qubit may be modeled using the interaction Hamiltonian

$$H_{env}^{ad} = \frac{\chi}{2}\left[\sigma_-^a\sigma_+^{e_a} + \sigma_+^a\sigma_-^{e_a}\right], \qquad (77)$$

where $\sigma_\pm = \sigma_x \pm i\sigma_y$ are spin raising and lowering operators. This is similar to the usual quantum-optical beam-splitter Hamiltonian $H_{BS} = \chi(a^\dagger b + b^\dagger a)$, and results in the same kind of coupling for our purposes. Following the same procedure as above, we find the amplitude damped rotation and CN gates to have fidelities

$$\mathcal{F}_R^{ad} = 1 - 1.80\lambda \quad (78)$$
$$\mathcal{F}_{CN}^{ad} = 1 - 2.20\lambda. \quad (79)$$

In Sec. III D we will find that these results are useful – they may be composed to establish an upper bound on the fidelity of an entire circuit with multiple R and CN gates.

## III. EXTENDING QUBIT LIFETIMES

We now apply the theories developed in the first part of this paper to describe the process of quantum error correction, and how it may be utilized to artificially extend the length of time for which a qubit can be maintained in a coherent superposition, despite the effects of decoherence. As an example, we concentrate on the correcting the errors due to phase damping, and present a detailed analysis of a three-bit circuit which can perfectly correct a single error. We calculate the fidelity of the perfect circuit, then apply our theory of noisy logic operators to calculate the fidelity of a circuit with imperfect logic gates.

We show that despite the imperfections, in a certain regime, an *effective decoherence rate* smaller than the actual decoherence rate can be achieved for the encoded "persistent qubit." Combining the simulation results with the theoretical predictions for the fidelities of noisy gates, we construct a general error model which predicts an upper limit on the acceptable decoherence per timestep figure of merit, $\lambda_{crit}$, which is a function of the theoretical effectiveness of the error correction code, and the complexity of the circuit required.

### A. Quantum Error Correction

An important conclusion from Sec. II C was that the effects of phase damping on a quantum superposition state between $|0\rangle$ and $|1\rangle$ can be understood as a classical, stochastic bit-flipping process. This realization immediately establishes a connection to classical coding theory, as was first recognized by Shor [7]. For example, we may use Eqs.(36-37) to interpret the errors introduced by phase damping in terms of a binary symmetric channel, as shown in Fig. 3. In the classical model, $|0\rangle$ and $|1\rangle$ are independent signals, and crossover errors in the noisy channel cause the signals to be interchanged. The quantum case is more subtle, because information is kept in the coherence between $|0\rangle$ and $|1\rangle$ and this will be lost by classical error correction schemes (see [8] for more discussion).

Decoherence is a quantum process which leads to destruction of quantum information. Although we have an analogy between noisy quantum processes and classical information theory, it is incomplete – we cannot naively utilize classical codes to preserve quantum information. This is possible only if codes are devised which leave a portion of the Hilbert space untouched. In other words, the codes must have a degeneracy. Fortunately, a solution using codes and their duals has now been worked out in great detail by Shor, Steane, and others. We present here a example using three qubits and discuss the physics in detail.

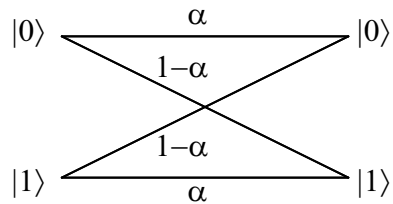

FIG. 3. Binary symmetric channel, with crossover error probability $1 - \alpha$. The input is on the left.

Suppose that we start with the state

$$|\psi_{in}\rangle = a\,|\text{---}\rangle + b\,|\text{+++}\rangle, \quad (80)$$

which results from encoding the qubit of Eq.(1) using two controlled-not (CN) gates and three single bit rotations. Note that this is a *single qubit of information* encoded using three physical qubits. Using Eq.(44), we find that phase damping causes us to obtain the mixed state output

$$\begin{aligned}
|\psi_{out}\rangle = &\, \alpha^{3/2} \left[ a\,|\text{---}\rangle + b\,|\text{+++}\rangle \right] \\
&\oplus \alpha\sqrt{1-\alpha} \begin{bmatrix} a\,|\text{--+}\rangle + b\,|\text{++-}\rangle \\ a\,|\text{-+-}\rangle + b\,|\text{+-+}\rangle \\ a\,|\text{+--}\rangle + b\,|\text{-++}\rangle \end{bmatrix} \\
&\oplus \sqrt{\alpha}(1-\alpha) \begin{bmatrix} a\,|\text{-++}\rangle + b\,|\text{+--}\rangle \\ a\,|\text{+-+}\rangle + b\,|\text{-+-}\rangle \\ a\,|\text{++-}\rangle + b\,|\text{--+}\rangle \end{bmatrix} \\
&\oplus (1-\alpha)^{3/2} \left[ a\,|\text{+++}\rangle + b\,|\text{---}\rangle \right]. \quad (81)
\end{aligned}$$

The vertically grouped terms should be understood to each be separate terms with the same prefactor; they have been placed together for emphasis of the following point: there are four distinct possibilities. Either no error occurs, or one, two, or three bits are flipped, in decreasing order of probability. Let us concentrate on the first two groups, which have the highest probability.

In classical error correction, a syndrome is calculated for each received word which identifies the error (if any). For quantum error correction, two orthogonal states must have the same syndrome, such that the calculation leaves some quantum coherence intact. Here, we calculate the two-bit syndrome given in Table I for which bit complements are degenerate (e.g., both +++ and --- have the same syndrome, 00). The result tells us how the error may be corrected, as long as only one bit flip has occurred. Such a scheme to detect and correct errors is depicted schematically in Fig. 4. Physically, we may consider the two additional qubits introduced in Eq.(80) as "probes" which allow us to detect what the environment does to our original single qubit. This is made possible by initially correlating the probes with our system, and then measuring the probes in the correct basis after decoherence. Assuming that only one error occurs, the measurement indicates which of the effects of Eqs.(11-12) has occurred, allowing us to correct the state of our qubit.

| Input   | Syndrome | Action         |
|---------|----------|----------------|
| --- +++ | 00       | no error       |
| --+ ++- | 01       | flip first bit |
| -+- +-+ | 10       | flip second bit|
| +-- -++ | 11       | flip third bit |

TABLE I. Syndromes for the eight three-qubit states.

In reality, two and three bit errors may occur; when these probabilities are taken into account, we find that the output qubit has imperfect fidelity; specifically, after correcting for single bit errors, we get the output

$$|\psi'_{out}\rangle = \sqrt{\alpha^3 + \alpha^2(1-\alpha)}\left[a\left|---\right\rangle + b\left|+++\right\rangle\right]$$
$$\oplus \sqrt{\alpha(1-\alpha)^2 + (1-\alpha)^3}\left[a\left|+++\right\rangle + b\left|---\right\rangle\right] \quad (82)$$

which has the fidelity

$$\mathcal{F} = \min |\langle\psi_{in}|\psi'_{out}\rangle|^2 \quad (83)$$
$$\geq \frac{1}{2} - \frac{e^{-3\lambda}}{4} + \frac{3e^{-\lambda}}{4} \approx 1 - \frac{3\lambda^2}{4} + \mathcal{O}(\lambda^3). \quad (84)$$

This may be compared with the result for no error correction, Eq.(6), in which $\mathcal{F}$ decreases as $\lambda$ instead of as $\lambda^2$. The improvement granted by this error correction scheme promises the possibility for extending the period for which a qubit can be preserved with high fidelity. We study this in detail next.

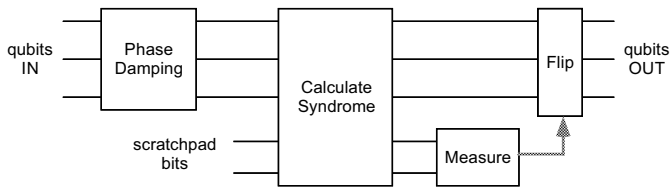

FIG. 4. Scheme for detecting and correcting single bit-flip errors occurring to the state $a\left|---\right\rangle + b\left|+++\right\rangle$.

## B. A Three Bit Circuit

Frequent error correction ensures that the probability of uncorrectable errors occurring remains small, and thus one might hope that a single qubit could thus be stored with high fidelity indefinitely. For example, a loop could be used to store an encoded qubit which circulates periodically through a circuit which detects and corrects errors (Fig. 5). Ancillary scratchpad qubits prepared in a definite initial state would be used for the syndrome calculation and then discarded, and in this manner a constant entropy flow can be maintained (order flows in through the ancilla, and out into the environment via dephasing of the stored qubit).

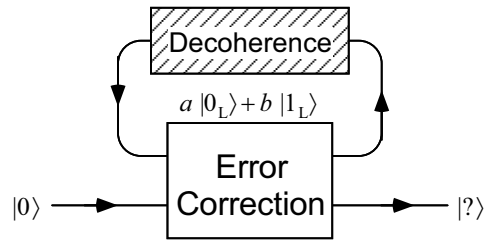

FIG. 5. Schematic of a system for preserving a quantum bit using periodic quantum error correction. $|0_L\rangle$ and $|1_L\rangle$ are multiple-qubit states which encode logical zero and one. The lower wire carries scratchpad qubits which are discarded (or reset) after the calculation.

For example, a circuit which implements the three-bit scheme described previously is shown in Fig. 6. Five timesteps are required, in which as many operations are performed simultaneously as possible. Broadly speaking, during the first two timesteps, the qubit is decoded (by converting from the Bell basis back into the computational basis) and the syndrome is calculated. The syndrome result is measured in the third timestep, and a classical computer corrects any detected error by performing a single bit flip. During the final two timesteps, the qubit is re-encoded.

An example will serve to explain the ideal operation of this machine: suppose that a single bit error due to phase damping occurs, transforming the qubit of Eq.(80) into the state $|\psi_0\rangle = a\left|+--\right\rangle + b\left|-++\right\rangle$ which is input to the circuit. In the first step the qubits are decoded:

$$|\psi_1\rangle = a\left|100\right\rangle + b\left|011\right\rangle \quad (85)$$

since $R$ simply rotates single bits from the Bell basis. After the second timestep we have

$$|\psi_2\rangle = a\left|111\right\rangle + b\left|011\right\rangle. \quad (86)$$

The two CN gates cause the syndrome to be calculated and left in the second and third qubits. This clever "in-place" calculation originally appeared in [8]. Note how the first qubit is left un-entangled with the syndrome bits. The measurement in the third timestep thus leaves us with

$$a\left|1\right\rangle + b\left|0\right\rangle, \quad (87)$$

which is the inverse of the original state, Eq(11). In fact, the syndrome measurement result 11 tells us that the third bit (from the right; the "most significant bit" in the string) is flipped, so that in the we may correct the error by applying a $\sigma_x$ operation, and arrive at

$$\left|\psi_3\right\rangle = a\left|000\right\rangle + b\left|100\right\rangle, \quad (88)$$

Next, coding is performed by the reverse process: the three bits are entangled by the two CN gates giving

$$\left|\psi_4\right\rangle = a\left|000\right\rangle + b\left|111\right\rangle, \quad (89)$$

and finally, rotations put the qubits into Bell states,

$$\left|\psi_5\right\rangle = a\left|---\right\rangle + b\left|+++\right\rangle, \quad (90)$$

giving a perfect qubit.

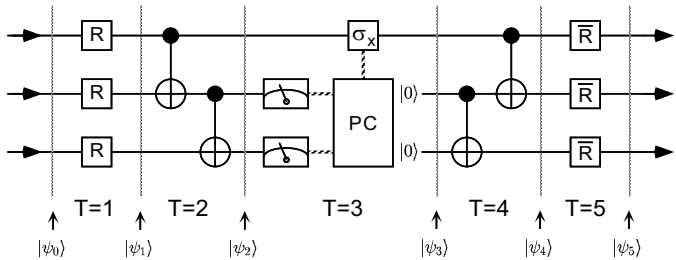

FIG. 6. Quantum circuit for performing error correction for a three-bit code. $R$ is a $\pi/2$ rotation which flips between the computational and the Bell basis, and $\overline{R}$ is the inverse operation. Vertical lines denote a controlled not, with the control line distinguished by the solid dot and the signal by a cross. Meters denote ideal measurements in the $\{\left|0\right\rangle, \left|1\right\rangle\}$ basis. PC is a classical computer, and $\sigma_x$ is the Pauli spin matrix operator.

Here, we have seen how the syndrome calculation and measurement leaves the quantum coherence of the encoded qubit intact. Furthermore, knowledge of the syndrome result indicates if the qubit is intact, or if any one of three single bit errors has occurred. When only one bit error occurs, the output will always be $a\left|---\right\rangle + b\left|+++\right\rangle$, but if more than one bit error occurs, we get $a\left|+++\right\rangle + b\left|---\right\rangle$, which is the wrong result. However, this occurs with smaller probability. The result is that by virtue of the error correction, the fidelity of a single (pure state) qubit encoded as specified will not be $\approx 1 - 6\lambda/2$, after one cycle through the circuit, but rather, $1 - 3\lambda^2/4$. Because this encoded qubit effectively *decoheres slower* than a single isolated qubit, we refer to it as a *persistent qubit*. Physically, it may be envisioned as a metastable collective state which has a lifetime that is kept artificially long by an active measurement and correction process.

### C. Processing Errors

Unfortunately, our quantum bit memory does not operate ideally, due to errors which occur during the finite time required for decoding, error detection, correction, and re-coding of the persistent qubit. For example, energy could be lost from the system during operation of the logic gates, the syndrome bits could be imperfectly measured, or phase randomization could occur. This is a serious issue, because it is not clear that the gains achieved by error correction may be realized in view of the additional errors incurred during the required processing!

The key figure of merit is the amount of decoherence per timestep $\lambda$, because this determines how much error occurs during processing. For example, suppose that we have the Hamiltonian

$$H = g\left[c^\dagger c(a^\dagger b + b^\dagger a)\right]$$
$$+ \lambda_0\left[a^\dagger a(e_a^\dagger + e_a) + b^\dagger b(e_b^\dagger + e_b) + c^\dagger c(e_c^\dagger + e_c)\right] \quad (91)$$

where the first bracketed term is the Hamiltonian for a logic gate (the quantum optical Fredkin gate), and the second term describes coupling with the environment (phase damping). After time $\tau = \pi/g$ the logic gate completely switches, but simultaneously, decoherence of amount $\lambda = \lambda_0 \tau$ occurs. This causes errors in the switching, which lead to imperfect error correction. Current experimental quantum computer realizations suffer from large $\lambda$; how critically will this limit the feasibility of using quantum error correction to provide long term qubit storage?

To better understand this issue, we performed extensive numerical simulations of the three qubit circuit presented in the previous section. The input is chosen to be a superposition state

$$\left|\psi_{in}\right\rangle = a\left|0\right\rangle + b\left|1\right\rangle, \quad (92)$$

It fed into the $T = 4$ step of an *imperfect* version of the circuit in Fig. 6, which produces after 2 timesteps an imperfectly coded qubit described by the density matrix $\rho_{pq}$, which is calculated by applying two CN and three R gates to the input. $\rho_{pq}$ is stored for $M$ timesteps during which phase damping occurs, calculated using Eq.(48), with $\alpha = (1 + e^{-M\lambda})/2$. This result, $\rho'_{pq}$ is then fed into the decoding circuit, steps $T = 1$ through 3 of the circuit, giving the output $\rho_{out}$, calculated by applying three R and two CN gates to $\rho'_{pq}$. We evaluate the output of this single processing cycle by calculating

$$\mathcal{F}_{cycle} = \min_{a,b}\left[\left\langle\psi_{in}\right|\rho_{out}\left|\psi_{in}\right\rangle\right], \quad (93)$$

$\mathcal{F}_{cycle}$ may be compared against two benchmarks: the fidelity of an isolated single qubit after the same elapsed time using Eq.(6):

$$\mathcal{F}_{single} = \frac{1 + e^{-(M+5)\lambda}}{2}, \qquad (94)$$

and the qubit fidelity after one cycle of an ideal circuit,

$$\mathcal{F}_{ideal} = \frac{2 - e^{-3M\lambda} + 3e^{-M\lambda}}{4}, \qquad (95)$$

from replacing $\lambda$ with $M\lambda$ in Eq.(84).

As a first attempt to qualitatively evaluate the impact of possible processing errors, we assume that each bit independently undergoes phase damping of an amount $e^{-\lambda}$ after the logic operation in each timestep. This is calculated by applying Eq.(48), with $\alpha = (1+e^{-\lambda})/2$. We assume that the decoherence rate (coupling strength to the environment) during processing is the same as during storage. This is not necessarily true, of course, but we shall amend this problem in the next section. The results from this naive model, shown in Fig. 7, are promising in that there exists a regime, in which despite the errors, we find that the fidelity of the persistent qubit is better than for a single qubit, i.e., $\mathcal{F}_{cycle} > \mathcal{F}_{single}$.

with the noisy rotation operators of Eqs.(69-70), and the noisy controlled-not operators of Eq.(75). For example, the output of the first CN gate in the fourth timestep is given by

$$\rho_{4a} = \sum_{k=\{00,01,10,11\}} C_k \rho_3 C_k^\dagger, \qquad (96)$$

where the density matrices describe the state of the three qubits used in the circuit.

Theoretically, this calculation is equivalent to a full density matrix calculation including all environmental degrees of freedom. However, in practice, such a calculation would involve 64×64 sized matrices (using one additional qubit to model the environment for each qubit in the perfect circuit) – and for larger systems this becomes impractical. Use of the noisy operator formalism allowed us to use only 8×8 matrices in the entire calculation, with the cost of two extra matrix multiplications for the R gate, and four for the CN gate. The result is shown in Fig. 7; a similar result for amplitude damping during processing is shown in Fig. 8.

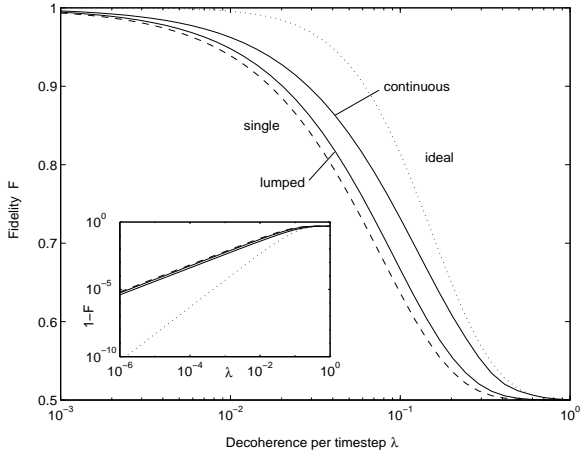

FIG. 7. Fidelity $\mathcal{F}_{cycle}$ of the persistent qubit after a single cycle through the imperfect error correction apparatus (solid line) – with "lumped" and continuous phase damping. For comparison, $\mathcal{F}_{single}$ (dashed line) and $\mathcal{F}_{ideal}$ (dotted line) are also shown. For small $\lambda$, $1 - \mathcal{F}_{cycle} \approx 5.50\lambda$ (lumped) and $1-\mathcal{F}_{cycle} \approx 3.92\lambda$ (continuous), as shown in the inset. $M = 8$.

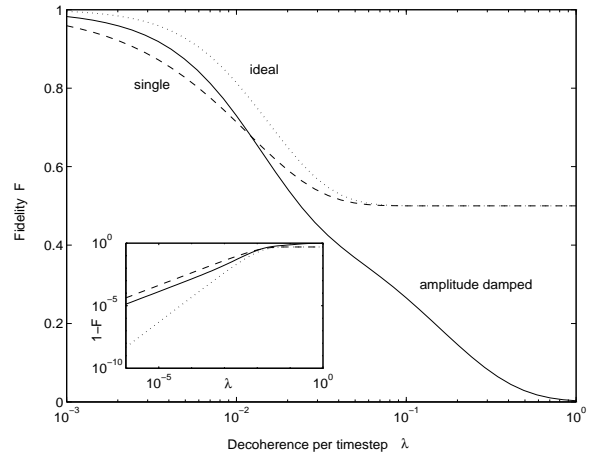

FIG. 8. Fidelity $\mathcal{F}_{cycle}$ of the persistent qubit after a single cycle through the imperfect error correction apparatus (solid line) – with continuous *amplitude* damping and $M = 80$. $\mathcal{F}_{single}$ (dashed line) and $\mathcal{F}_{ideal}$ (dotted line) are also shown. For small $\lambda$, $1 - \mathcal{F}_{cycle} \approx 13.5\lambda$.

However, although this naive "lumped decoherence" model, in which phase damping is artificially inserted after the logic operation in each timestep, is a practical way to evaluate the qualitative effect of decoherence, it unfortunately does not necessarily reflect reality. True decoherence occurs *simultaneously* with the logic operation, and can cause much more subtle errors by reducing not only the fidelity of single qubits but also the fidelity of the entanglement between qubits [25]. We evaluated the effect of phase damping during the logic operations by replacing the ideal R and CN operators in the circuit

We find that for this particular circuit, the lumped decoherence model actually overestimates the error, and despite the processing error, regimes exist for which the persistent qubit outperforms the single qubit. Furthermore, errors due to amplitude damping during processing result in a worse $\mathcal{F}_{cycle}$ than for phase damping. As discussed in Sec. II E, this may be because amplitude damping affects the diagonal as well as the off-diagonal elements of the density matrix. Thus, contrary to popular believe, relaxation may actually be more damaging to quantum computing than phase damping! Clearly, further work must be done to clarify this issue.

Now, the results from both of these simulations seem to indicate that the fidelity of the persistent qubit is only marginally better than for a single qubit. However, it must be kept in mind that this is true only after a single cycle through the circuit. The great advantage of error correction arises from its effect in limiting the *accumulation* of errors. Due to decoherence, the fidelity of a qubit decreases multiplicatively, such that a small amount of error grows exponentially with time. Thus, after time $t$, an isolated single qubit with no error correction has fidelity

$$\mathcal{F}_{single}(t) = \frac{1 + e^{-\lambda t}}{2} \approx 1 - \frac{\lambda}{2} t \qquad (97)$$

In contrast, with error correction, the error growth rate is slowed by correcting a fraction of error in each pass through the apparatus. We may model the fidelity of the result as

$$\mathcal{F}_{pqubit}(t) = \frac{1 + e^{-\lambda_{eff} t}}{2}, \qquad (98)$$

where the *effective decoherence rate* $\lambda_{eff}$ for the persistent qubit is

$$\lambda_{eff} = -\frac{1}{M+5} \log \left[ 2\mathcal{F}_{ideal} - 1 \right], \qquad (99)$$

that is, the logarithm of the error probability after a single correction step. For small time $t$ and decoherence $\lambda$, we find that the persistent qubit has fidelity

$$\mathcal{F}_{pqubit}(t) \approx 1 - \frac{3M^2 \lambda^2}{4(M+5)} t \qquad (100)$$

when errors occur only during storage, and not during processing. This model agrees quite well with our numerical simulation of a perfect circuit; however, as can be seen from Figs. 7-8, this expression must be modified. We shall next see how processing errors cause the fidelity of the persistent qubit to decrease not as $\lambda^2$, but rather, at best, as $\lambda^{3/2}$.

### D. General Error Model

Generally speaking, a persistent qubit is created by using a quantum error correction scheme to encode a single qubit of information using multiple qubits. The effects of decoherence during storage imposed by unwanted interactions of this many-body state with the environment can be detected and undone by the decoding circuit, as long as the error is not too severe. Furthermore, it is important not to introduce too many additional errors during coding and decoding. These requirements can be modeled in a general way, giving a result which is applicable for an arbitrary scheme, as we show here.

Two competing processes happen to a persistent qubit: on one hand, as the storage time $M$ is increased, the probability of unrecoverable (multiple-bit) errors happening increases, which is bad. However, also as $M$ increases, the fraction of time spent out of the noisy circuit decreases, which is good. Thus, an optimal value for the qubit storage time $M_{opt}$ exists.

We may calculate $M_{opt}$ using the following general model, good for small decoherence $\lambda$. Motivated by the result in Eq.(95), let us assume that a generalized $N$-bit coded persistent qubit has fidelity

$$\mathcal{F}_{storage} = 1 - \alpha M^2 \lambda^2, \qquad (101)$$

where $\alpha$ is the probability of an uncorrectable error occurring despite having an ideal circuit. This is proportional to $\lambda^2$ because we assume that only single bit errors are corrected for; for a better scheme, the exponent of $\lambda$ should be increased. Note that we do not assume what kind of error the circuit corrects for − it may be phase damping, or even any single bit error. The fidelity of the circuit is modeled as

$$\mathcal{F}_{circuit} = 1 - \beta \lambda + \beta' \lambda^2 \qquad (102)$$

after the $N$ timesteps required for the circuit to operate. $\beta$ is thus the probability of an error occurring in an imperfect circuit. Phenomenologically, any systematic error, for example, errors in physical implementation of the computing system, including possible measurement errors, could also be included in this parameter. For completeness, we also include a higher order term, $\beta'$, but it will not be relevant to the first-order solutions we obtain below.

Now, in our system, the qubit is stored for $M$ timesteps, then processed by the circuit. Each of these cycles leaves a qubit with fidelity

$$\mathcal{F}_{cycle} = \mathcal{F}_{circuit} \mathcal{F}_{storage} \qquad (103)$$

Since each cycle requires $M+N$ timesteps, we may define an effective decoherence rate

$$\lambda_{eff} = -\frac{1}{M+N} \log \left[ 2\mathcal{F}_{cycle} - 1 \right]. \qquad (104)$$

Working this out, we find that $\lambda_{eff}$ is proportional to $\lambda$ for small decoherence, and dependent on the storage time $M$. The numerical result, shown in Fig. 9, indicates the existence of an optimal value $M_{opt}$ at which $\lambda_{eff}$ is minimal, as expected. As an aside, it is interesting to note that the input state for which the fidelity is minimum is $(|0\rangle + |1\rangle)/\sqrt{2}$ for $M < M_{opt}$, and $|0\rangle$ for $M > M_{opt}$. That is because in the first regime, the decoherence of the imperfect circuit dominates, while in the second case, the decoherence during storage dominates. This switch is part of the reason why the transition appears to be so abrupt.

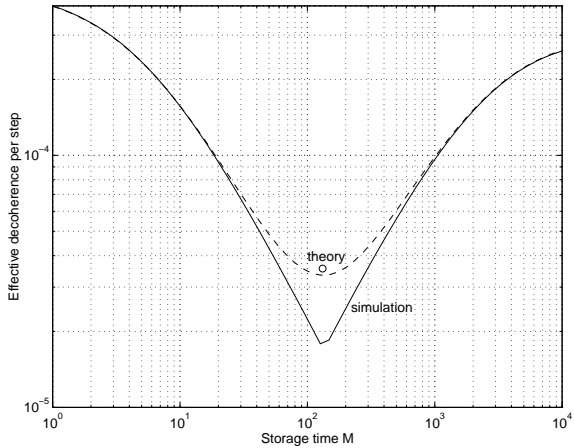

FIG. 9. $\lambda_{eff}$ as a function of storage time $M$, simulation (solid) and theoretical (dashed) results, calculated for our three-bit circuit model, and $\lambda = 3 \times 10^{-4}$. The circle is at $(M_{opt}, \lambda_{opt})$.

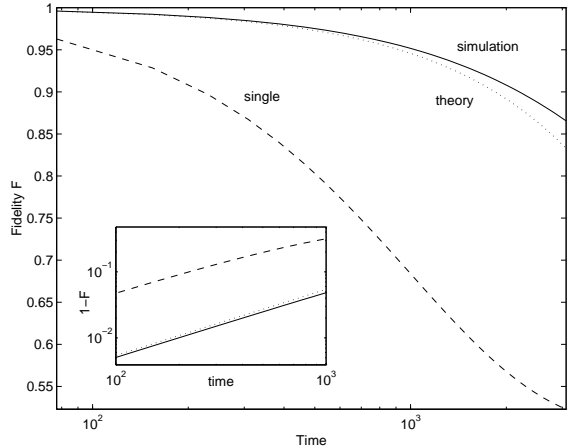

FIG. 10. Fidelity $\mathcal{F}_{cycle}(t)$ of the three-bit persistent qubit as a function of time, simulation result assuming processing errors due to continuous phase damping (solid); theoretical optimum $\mathcal{F}'_{opt}(t)$ (dotted); and single-bit case $\mathcal{F}_{single}(t)$ (dashed), all calculated for $\lambda = 10^{-3}$ and $M_{opt} = 72.1$.

The minimum $\lambda_{eff}$ is obtained when

$$M_{opt} \approx \sqrt{\frac{\beta}{\alpha\lambda}} - N, \qquad (105)$$

to lowest order in $\lambda$. As $\beta \to 0$, then no processing errors occur, and we find that the optimal storage time $M_{opt} \to 0$ as expected (it cannot be negative). Note that $M_{opt} \sim 1/\sqrt{\lambda}$. Plugging back into $\lambda_{eff}$, we find that the minimum achievable effective decoherence rate is theoretically

$$\lambda_{opt} \approx 4\sqrt{\alpha\beta}\lambda^{3/2} \qquad (106)$$

and the corresponding optimal persistent qubit fidelity is

$$\mathcal{F}_{opt}(t) \approx 1 - 2\sqrt{\alpha\beta}\lambda^{3/2}t. \qquad (107)$$

This result describes the fidelity of a persistent qubit with decoherence during processing, and should be compared with Eq.(100). For small $\lambda$, this is still better than $\mathcal{F}_{single} \sim 1 - \mathcal{O}(\lambda)t$, but is worse than $\mathcal{F}_{pqubit} \sim 1 - \mathcal{O}(\lambda^2)t$ because of the noisy circuit.

Numerically, we find in our specific three-bit circuit that the multiplicative accumulation of errors due to periodic correction results in a minimum effective decoherence rate $\lambda'_{opt}$ which is a factor of two smaller than $\lambda_{opt}$ for small $\lambda$. Taking this into account, we find that for the parameter values of $\alpha = 0.75$, $\beta = 3.92$, and $N = 5$, our simulation results are well modeled by $M_{opt} \approx 2.28/\sqrt{\lambda}$, $\lambda'_{opt} \approx 3.43\lambda^{3/2}$, and $\mathcal{F}'_{opt}(t) \approx 1.71\lambda^{3/2}t$, as shown in Fig. 10.

At some point, the effective decoherence rate of the persistent qubit is no longer less than the decoherence rate of an isolated single qubit, i.e. $\lambda_{eff} \approx \lambda$. This happens when the circuit noise completely overwhelms the effectiveness of the code in correcting for errors. Analytically an order of magnitude estimate for an upper bound on $\lambda$ is given by $\lambda_{crit} \approx 1/4\beta\sqrt{\alpha}$ (to lowest order in $\alpha$). Here, $\alpha$ is the probability that an uncorrectable error occurs; by improving the coding scheme, this can be reduced. However, doing so may increase the complexity of the circuit, and thus increase $\beta$, the probability of a processing error occurring. At worst, $\beta$ is proportional to the total number of gates in the circuit; this is likely to be the case for persistent qubit circuits, in which entanglement probably involves all qubits. From an algorithmic standpoint, there are undoubtedly optimal configurations, but that is beyond the scope of this paper.

The formalism presented in this section may be applied to any general quantum bit memory system. In fact, the noisy gate fidelities derived in Sec. II F may be used immediately to place an upper bound on the performance of an arbitrary circuit. For example, if we assume that phase damping occurs during the processing of our three-bit circuit, we may estimate that $\beta < 6 \times 0.40 + 4 \times 0.86 = 5.84$, from counting six R gates and four CN's in the circuit. This differs from the $\beta = 3.92$ arrived at from the simulation because the state which minimizes the fidelity of the circuit is different from that for just the single gate, and also because not all gates must work perfectly for the circuit to behave correctly. However, in general, the $\alpha$, $\beta$ model is useful in that the end result of various physical effects may be estimated immediately. For example, different decoherence rates during storage and processing can be accounted for by adjusting these parameters.

Thus, knowledge of the circuit structure and the individual gate fidelities can be used to bound the performance of a persistent qubit circuit. From $\alpha$ and $\beta$, a critical value which establishes a minimum required decoherence per timestep threshold can be estimated. If the decoherence in the experimental system is worse than $\lambda_{crit}$ then the circuit will be only marginally viable. On the other hand, if $\lambda < \lambda_{crit}$ is achieved, then good results can be expected. The example of Fig. 10 shows that for a circuit with decoherence happening during processing, the lifetime (defined as time until the fidelity falls below 0.95) of a persistent qubit can be an order of magnitude longer than for a single isolated qubit.

## IV. CONCLUSION

We have presented a new formalism for modeling the decoherence of quantum bits, based a general linear operator description of quantum mechanics. From this, we derived simple matrix operators to model phase damping, Eqs.(36-37) and Eq.(44), and amplitude damping, Eqs.(54-56) of idle "memory" qubits. We also derived "noisy logic gate operators" for qubits undergoing a $\pi/4$ rotation, Eqs.(69-70), and the conditional dynamics of a controlled-NOT, Eq.(75). These operators allow immediate evaluation of logic gate fidelities, and because they act only in the Hilbert space of the system, they also allow simulation of decoherence without including extra states for the environment.

Our application of these results to study a quantum error correction system indicate that despite decoherence during processing, when the decoherence per timestep is smaller than $\lambda_{crit}$, the circuit can still be effective. In fact, our model predicts that for a circuit with error probability $\beta\lambda$ due to decoherence during processing, and $\alpha\lambda^2$ due to the imperfection of the algorithm, a persistent qubit can be constructed with fidelity $1-2\sqrt{\alpha\beta}\lambda^{3/2}t$. Physically, the persistent qubit may be understood to be a metastable collective state whose lifetime is artificially prolonged by repeated decoding, measurement, and encoding.

A persistent qubit would be much simpler to implement than any of the elementary quantum algorithms proposed thus far, and in particular, much easier to accomplish than factoring. Furthermore, it would be a useful step towards an eventual goal of indefinite storage of quantum information. Extension of our analysis to storage of physically separated, entangled qubits is straightforward, and would be relevant for realizations of concepts such as quantum money and quantum teleportation [26]. We thus suggest as an alternative and, we believe, more practical first step − the utilization of quantum error correction techniques to implement a long lifetime single-qubit memory.

Possible physical implementations include single photon quantum bits [21,5] or single ions in an electromagnetic trap [27,4]. For the latter case, our analysis could be extended to calculate noisy counterparts to Cirac and Zoller's $U$ and $V$ operators, in a manner similar to [15]. Together with a realistic estimate of systematic errors expected in an experimental implementation, this should result in $\alpha$ and $\beta$ parameters which may be used to evaluate the performance of an ion-trap persistent qubit.

ILC gratefully acknowledges the support of the Fannie and John Hertz Foundation.